\documentclass[useAMS,usenatbib,referee]{mn2e}

\usepackage{graphicx}
\usepackage{epsf}
\usepackage{fleqn}

\title{The true nature of CSL-1.}
\author[Sazhin M. et al.]
{Sazhin M.V.$^1$,  Capaccioli M.$^{2,3}$,  Longo G.$^{3,4}$,\\
\newauthor
Paolillo M.$^{3,4}$, Khovanskaya O.S.$^{1}$,\\
\newauthor
Grogin N.A.$^{6}$, Schreier, E.J.$^{7}$,  Covone, G.$^{5}$\\
 1 - Sternberg Astronomical Institute, Moscow State University,University pr. 13, Moscow, RUSSIA\\
 2 - VSTceN at the OAC, INAF - Osservatorio Astronomico di Capodimonte, via Moiariello 16, 80131, Napoli, ITALY\\
 3 - Department of Physical Sciences, University of Napoli Federico II, via Cinthia 9, 80126 Napoli, ITALY\\
 4 - INFN - Napoli Unit, Dept. of Physical Sciences, via Cinthia 9, 80126, Napoli, ITALY \\
 5 - INAF - Osservatorio Astronomico di Capodimonte, via Moiariello 16, 80131, Napoli, ITALY\\
 6 - Dep. of Physics and Astronomy, The Johns Hopkins University, 3400 North Charles Street, Baltimore, MD 21218, USA\\
 7 - Associated Universities Inc., Suite 730, 1400 16th St., Washington, DC 20036, USA}
\date{Accepted ;
      Received ;
      in original form }

\pagerange{\pageref{firstpage}--\pageref{lastpage}} \pubyear{2005}

\begin{document}

\maketitle

\label{firstpage}

\begin{abstract}
On January 12 of 2006, the Hubble Space Telescope (HST) observed the peculiar double extragalactic object CSL-1, suspected to be the result of gravitational lensing by a cosmic string. 
The high resolution image shows that the object is actually a pair of interacting giant elliptical galaxies. 
In spite of the weird similarities of the energy and light distributions and of the radial velocities of the two components, CSL-1 is not the lensing of an elliptical galaxy by a cosmic string.
\end{abstract}


\section{Pre-HST observations of CSL-1}
The Capodimonte-Sternberg-Lens candidate n. 1 (CSL-1; $\alpha_{2000}=12^h$ $23^m$ $30\fs5$, $\delta_{2000}=-12\degr$ $38\arcmin$ $57\farcs0$), a peculiar extragalactic object observed in the OACDF by \citet{csl1}, hereafter Paper I, appears as a double source projected onto a low density field. 
The two components of the source are 1.9 arcsec apart, well resolved (i.e. extended) and, when observed from the ground, show roundish and identical shapes. 
Low resolution spectra look identical with a high confidential level (c.l.) and both give the same redshift of $0.46\pm 0.008$. 
Photometry matches the properties of giant elliptical galaxies. 
Additional medium-high resolution observations carried on with FORS1 at VLT in March 2005 confirmed the similarity of the spectra of the two objects at a 98\% c.l. \citet{saz05} and 
showed that the differential radial velocity is compatible with zero $\pm 20$ km s$^{-1}$.

In view of all that, already in Paper I we argued that CSL-1 could be either {\it i)} the chance alignment of two very similar giant E galaxies at the same redshift, or {\it ii)} a gravitational lensing phenomenon.
In this second case detailed modelling shows that the properties of CSL-1 are compatible only with the lensing of an E galaxy by a cosmic string. 

Cosmic strings were first introduced by \citet{kib76} and extensively discussed in the literature (cf. \citet{zel} and \citet{vil}).
Recent work has also shown the relevance of cosmic strings for both
fundamental physics and cosmology (cf. \citet{dav05}) and has shown that cosmic strings could be observable, in principle, via several effects, the most important being gravitational lensing, as we extensively discussed in the above quoted papers.

\begin{figure*}
\begin{center}\label{HST1}
\includegraphics[height=5.5cm]{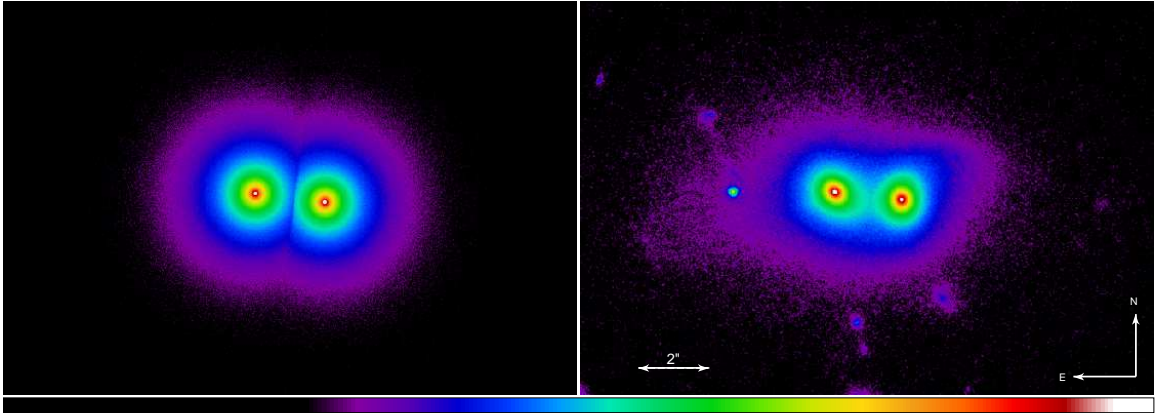}
\end{center}
\caption{Left panel: numerical simulation of the image of an E galaxy 
with a de Vaucouleurs [$r^{1/4}$] light profile, as it would appear if 
lensed by a cosmic string (the situation roughly corresponds
to that of CSL-1 and the noise level is that expected in our HST observations.). Right panel: the image region surrounding CSL-1 as observed by HST. A logarithmic colorscale has been adopted to enhance the  morphological details.}
\label{string}
\end{figure*}

Is CSL-1 a cosmic string? In Paper I, we suggested an {\it experimentum crucis} \citep{csl1}: the detection of the sharp edges
at faint light levels which are expected to be generated by a cosmic string (Fig.\ref{string}). 
This test requires high angular resolution and deep observations to be performed with the Hubble Space Telescope (HST). 
In what follows we shortly summarise the most immediate outcomes of these observations.

\section{The data}
The observations were performed on January 11 2006 with the ACS camera. CSL-1 was observed for 6 HST orbit in the F814W band (comparable to Johnson-Cousins I-band) yelding an effective exposure time of $\sim 14000$ seconds.
The observations were performed adopting a 1/3 pixel dither pattern, to allow sub-pixel sampling of the HST PSF and cosmic ray rejection. All 6 orbits were combined through the Multidrizzle software \citep{Koek02} using a 1/2 pixel (0.025 arcsec/pixel) resampling pattern. In what follows we shortly summarize the main results. 

\section{The HST images}
Figure \ref{HST1} shows the region surrounding CSL-1 as observed by the 
ACS Wide Field Camera on board of HST. The faint isophotes of the two components have different shapes, which is incompatible with CSL-1 being a cosmic string, and the expected sharp edges are not present. 
In the cosmic string scenario all morphological features of the source falling inside the deficit angle, would be mirrored on the opposite side of the string. However in the HST image 
we do not see such mirroring effect for the two components, nor for any other faint feature which, would have fallen inside the deficit angle of the string and
should have been duplicated, e.g. the faint sources on the southern side of CSL-1 visible in the right panel of Figure \ref{HST1}.

Another test, proposed and discussed by several
authors \citep{string1, string2, string5, saz04}, takes into account the fact that 
the alignment of the background object (a galaxy) inside
the deficit angle of the string is a stochastic process
determined by the area of the lensing strip and by the
surface density distribution of the extragalactic objects
which are laying behind the string. All objects falling
inside the narrow strip defined by the deficit angle
computed along the string pattern should be affected and, the deeper the image, the
larger the number of gravitationally lensed images to be expected. 

\begin{figure*}
\begin{center}
\includegraphics[width=15cm]{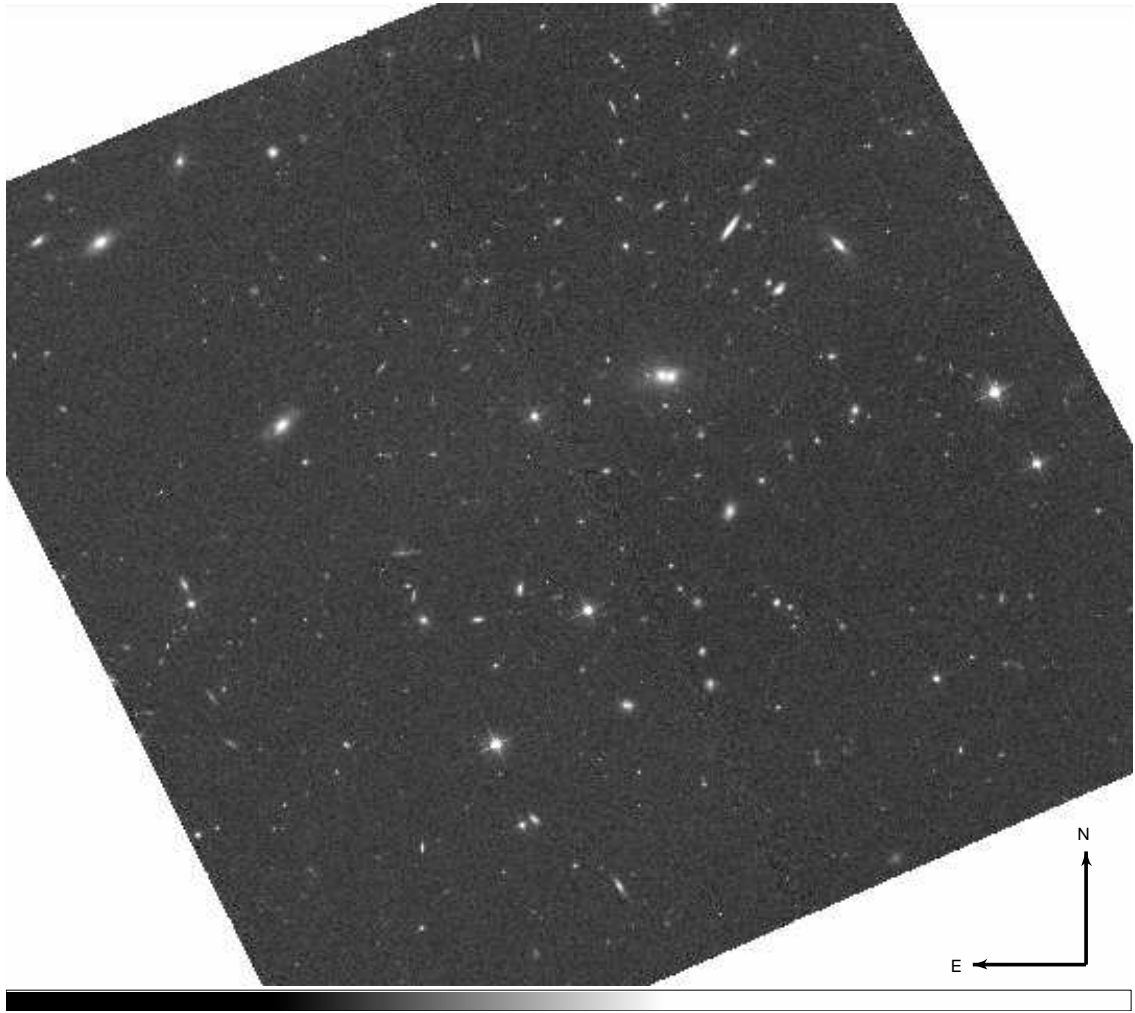}
\end{center}
\caption{The ACS field observed by HST. CSL-1 is the double object slightly off-centered.
The image is the composite of the 6 HST orbits.}
\label{milky}
\end{figure*}

In Fig.\ref{milky} we present the image of the whole field, produced by stacking the 6 HST orbits. Our careful examination of the source population around CSL-1 reveals no overdensity of galaxy pairs with separations $< 2"$. The lack of such galaxy pairs in our image (aside from CSL-1) therefore leads to a strong rejection of the cosmic string hypothesis.

\begin{figure*}
\begin{center}
\includegraphics[width=12.0cm]{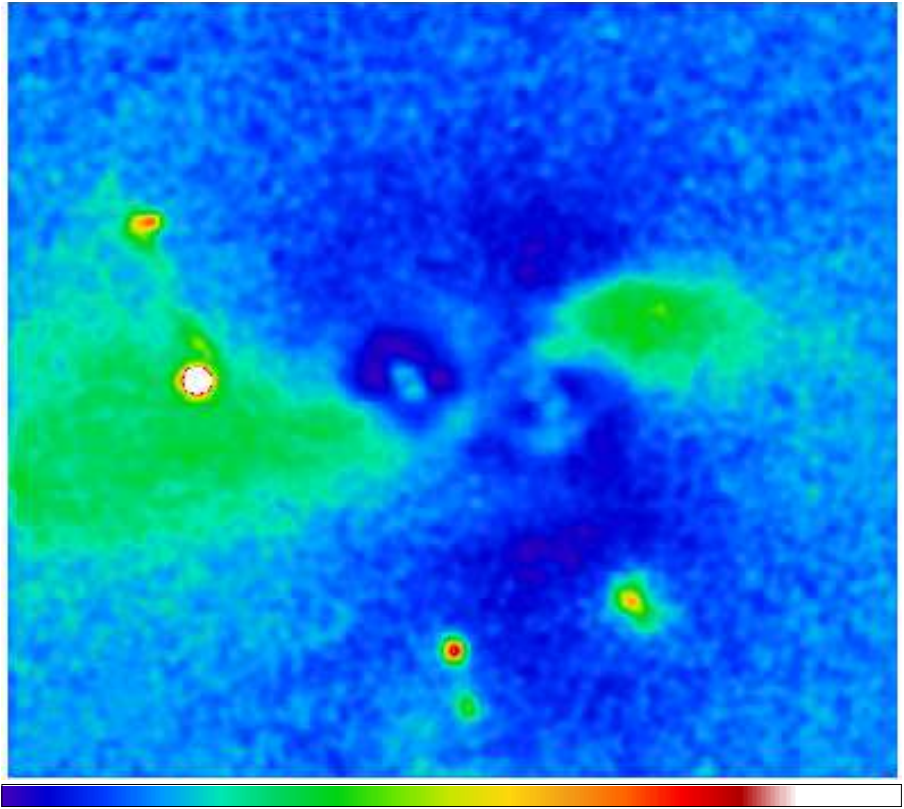}
\end{center}
\caption{The normalized residuals (residuals/model) obtained by subtracting from the HST images a model consisting of two de Vaucouleurs light profiles. }
\label{res}
\end{figure*}

For completeness we also fit the two objects with two de Vaucouleurs $r^{1/4}$ light profiles and subtract the model from the original data. The residual image, presented in Fig.\ref{res}, clearly shows the presence of warped structures in the CSL-1 outskirts, probably tidal tails due to the interaction between the two galaxies.

As a result of our observation, we have therefore to conclude that CSL-1 is not the lensing of an elliptical galaxy by a cosmic string.
\medskip

M.V. Sazhin acknowledges the INAF-VSTceN for hospitality and financial support, and the financial 
support of RFFI grant 04-02-17288. 
O.S. Khovanskaya acknowledges the Department of Physical Sciences of the University of Naples Federico II 
and the INFN-Napoli for financial support, as well as the financial support of the grants: 
President of RF "YS-1418.2005.2" and of MSU for YS. 
This research was partly funded by the Italian Ministry for Instruction, University and Research, 
through a COFIN-2004 grant no. 2004020323.

\label{lastpage}

\end{document}